\documentstyle[preprint,revtex,eqsecnum]{aps}

\begin{document}
\draft
\preprint{}

\begin{title}
                 Boson-fermion mappings for odd systems \\
                        from supercoherent states
\end{title}

\author{
        J. Dobaczewski,\cite{Warsaw} F. G. Scholtz, and
        H. B. Geyer
        }

\begin{instit}
                    Institute of Theoretical Physics,
                     University of Stellenbosch, \\
                    Stellenbosch 7600, South Africa
\end{instit}

\begin{abstract}
We extend the formalism whereby boson mappings can be
derived from generalized coherent states to boson-fermion
mappings for systems with an
odd number of fermions. This is accomplished by constructing
supercoherent states in terms of both complex and Grassmann
variables.
In addition to a known mapping for the full
so(2$N$+1) algebra, we also uncover some other formal
mappings, together with mappings relevant to collective
subspaces.
\end{abstract}

\narrowtext

\section{INTRODUCTION}
\label{sec1}

Phenomenological models of collective states in odd fermion
systems (mostly nuclei in the present context) usually
assume that these states can be approximated by states in
the product Hilbert space
   \begin{equation}\label{e1}
   {\cal H} = {\cal H}_{\rm even} \otimes {\cal H}_
              {\rm s.p.} ,
   \end{equation}
where ${\cal H}_{\rm even}$ denotes the Hilbert space of
collective states in the neighboring even-even system, and
${\cal H}_{\rm s.p.}$ the Hilbert space of
single-particle states.  The particle-plus-rotor model
\cite{BM75} constitutes a classical example of a model
constructed in such a way.  The Hilbert space ${\cal
H}_{\rm even}$ is constructed in this case as a model
space of a rotor without any explicit reference to a
microscopic description of states in an even-even core.
The Pauli correlations between the odd fermion and the
fermions comprising the even-even core are thus simply
neglected.  A similar approximation is also made in the
quasiparticle-plus-core model \cite{DF78}, where pairing
correlations are taken into account by considering in
${\cal H}_{\rm s.p.}$  quasiparticles instead of
particles, and in ${\cal H}_{\rm even}$ both neighboring
even-even cores.

In the phenomenological models of such odd fermion systems
the Hamiltonian is assumed to be of the form
   \begin{equation}\label{e2}
   \hat H = \hat H_{\rm even}
          + \hat H_{\rm s.p.} + \hat H_{\rm int} ,
   \end{equation}
where the three components describe the even-even core, the
single-particle states, and the interaction between them,
respectively.
Although the interaction mixes the eigenstates of $\hat
H_{\rm even} + \hat H_{\rm s.p.}$, it is usually
introduced to describe dynamical effects rather than
corrections induced by the neglect of Pauli correlations in
the basis states of the full $\cal H$.

It is worthwhile to recall here that even a
Slater-determinant wave function of an odd nucleus,
   \FL\begin{equation}\label{e3}
   \Psi(x_1\ldots x_A) = {\textstyle\frac{1}{\sqrt{A!}}}\sum_P(-1)^P
            \psi_{i_1}(x_1)\ldots\psi_{i_A}(x_A) ,
   \end{equation}
(where $P$ is a permutation of indices, $P(1,\ldots,A) =
i_1,\ldots,i_A$, with $(-1)^P$ its parity), cannot be
presented as a simple product of a single Slater
determinant of the core and of a single odd-fermion wave
function,
\widetext
   \begin{equation}\label{e4a}
   \Psi'(x_1\ldots x_A) = \left[{\textstyle\frac{1}{\sqrt{(A-1)!}}}
            \sum_{P'}(-1)^{P'}
            \psi_{i_1}(x_1)\ldots\psi_{i_{A-1}}(x_{A-1})
            \right]\times\psi_A(x_A) ,
   \end{equation}
(where $P'(1,\ldots,A-1) = i_1,\ldots,i_{A-1}$).  However,
the Slater determinant of Eq.{\ }(\ref{e3}) belongs to the
product Hilbert space (\ref{e1}), because it can be
presented as a linear combination of states (\ref{e4a}):
   \begin{equation}\label{e5}
   \Psi(x_1\ldots x_A) = \sum_{j=1}^A (-1)^{j+1}
            \left[{\textstyle\frac{1}{\sqrt{(A-1)!}}}
          \sum_{P'}(-1)^{P'}
            \psi_{j_1}(x_1)\ldots\psi_{j_{A-1}}(x_{A-1})
            \right]\times\psi_j(x_A) .
   \end{equation}
where the set of indices $j_1,\ldots,j_{A-1}$ comprises
$1,\ldots,A$ with the index $j$ excluded.
\narrowtext

In principle, we can therefore think about restoring Pauli
correlations by constructing an interaction $\hat
H_{\rm int}$ which would enforce or assure the mixing of
states (\ref{e4a}) in such a way as to obtain
states (\ref{e5}).  This task is virtually hopeless when the
even states are described by a model which does not
explicitly use fermion degrees of freedom.  In the present
study we consider and present relevant constructions when
the core states are described by bosons which result from a
rigorous boson (or boson-fermion) mapping. In this case it
becomes possible to address Pauli correlations between a
chosen core and surplus fermions in a systematic way.

A model for which such an analysis is of direct relevance
is the interacting boson-fermion model (IBFM) \cite{IvI91},
where Pauli correlations are at least partially accommodated
on the phenomenological level through an exchange term
which mimics the microscopic exchange of fermions between a
single fermion and a fermion pair. (There is microscopic
evidence that the fermion quadrupole pairing interaction may
be largely responsible for such an exchange term in the
IBFM ; see Ref.{\ }\cite{dKG88} and references therein.)

The general formalism of boson and boson-fermion mappings
or realizations of Lie algebras (from a nuclear physics
point of view) and their present status have recently been
reviewed extensively by Klein and Marshalek \cite{KM91}.
Amongst the open problems identified in that review is the
one discussed above, phrased in the terminology of
generalized quantized Bogoliubov-Valatin (QBV)
transformations, with a further systematization of such
transformations envisaged.  This refers precisely to an
approach where only some collective pair degrees of freedom
are earmarked for bosonization, while the remaining degrees
of freedom are to be treated as ideal fermions,
kinematically independent from the bosons.

QBV results which have so far been obtained pertain first
to the full {so(2$N$+1)} algebra where all fermion pairs are
bosonized and only states with at most one odd fermion
subsequently need to be considered in the product space
(\ref{e1}) (Ref.{\ }\cite{KM91} and references cited therein).
When bosons are associated with correlated fermion pairs
defined by some collective subalgebra (and the product
space (\ref{e1}) is naturally expected to contain states with
more than one odd fermion), QBV-type results have so far
only been obtained for a limited number of low rank
subalgebras, namely su(2) \cite{SM76,GH83}, su(3) \cite{KW90},
so(4) \cite{HK89} and so(5) \cite{KWGH91}.  Furthermore
these results have been obtained exclusively from algebraic
considerations, as opposed to derivation via coherent
states - the two main avenues which have been explored for
the mapping of even fermion systems.

In review \cite{KM91} algebraic considerations are mostly
stressed, although it is appreciated that the
coherent state approach has been instrumental in the
historical development, while also appealing for the
economy and elegance with which it leads to boson mappings
and the rigorous systematization of various mappings and
results.  As an example of the utility of the coherent
state approach, one may quote the natural appearance of the
${\cal R}$-projection which plays an important role in the
identification of spurious states as has been known for
some time \cite{Dob81b} and also vividly demonstrated
recently \cite{DGH91}.

It is therefore to be expected that a coherent state
approach to boson-fermion mappings of odd systems, and
ultimately generalizations of the QBV
transformation, will play an important complementary role
to present results and endeavors which exploit algebraic
methods.

In this paper we present the proper framework to address
the above program, namely introduce the appropriate
coherent states (supercoherent states) and report on some
first results.  We also comment briefly on some possible
further developments and hurdles which will have to be
overcome.  The organization of the paper is then as
follows: In Sec.{\ }\ref{sec2} we give a {\it r\'esum\'e} of
the background to generalized QBV mappings, stressing the
restrictions on states which are to be included in the
physical subspace of the ideal space.  We discuss
the distinction between ideal fermions and ideal
quasifermions which becomes important for a discussion of
properties of the ideal space.  Supercoherent states are
introduced in Sec.{\ }\ref{sec3a} for the {so(2$N$)} algebra.  We
also present there various similarity transformations and
define the mapping projected onto the space with at most
one ideal fermion.  In Sec.{\ }\ref{sec3b} we obtain
mappings induced by supercoherent states defined in the
collective space, and give some examples for this case
in Sec.{\ }\ref{sec4}.
Sec.{\ }\ref{sec5}
contains a discussion of what has been achieved
and where future effort should be directed to
obtain QBV-type mappings from supercoherent states
for collective spaces.

\section{QUANTIZED BOGOLIUBOV-VALATIN MAPPING \\
          AND STRUCTURE OF THE IDEAL SPACE}
\label{sec2}

We introduce the concept of a boson-fermion mapping and its
specialization to the quantized Bogoliubov-Valatin (QBV)
transformation (and possible generalizations) in the simple
setting of a single $j$-shell. Suppressing the index $j$,
we introduce fermion creation and annihilation operators
$a^\mu\equiv a_\mu^+$ and $a_\mu$, respectively,
where $\mu$ can take on $N=2j+1$ values.

The algebra of products
   \begin{equation}
   N_{\mu\nu} = a^\mu a_{\nu}, \label{eq:2}
   \end{equation}
   \begin{equation}
   A^{\mu\nu} = a^\mu a^{\nu} = (A_{\mu\nu})^{+},
   \label{eq:3}
   \end{equation}
generates the orthogonal algebra {so(2$N$)}.
If supplemented by all the commutators of single and
bifermion operators and the commutator of the single
fermion operators themselves, the corresponding algebra is
{so(2$N$+1)}.

We remark here that alternative to supplementing the
{so(2$N$)} algebra in the above fashion, one could of
course replace the commutators of single fermion operators
by the perhaps more natural anti-commutators, leading to an
equivalent algebraic structure which, however, will then
not be an algebra any more, but rather a superalgebra.
(This superalgebra has a rather simple structure as it can
be obtained by supplementing the algebra with its trivial
center, the identity.) To the extent that supercoherent
states will be used to induce the above algebraic (or
equivalently superalgebraic) structure in an ideal space,
these induced relations will typically hold on the whole
ideal space, whereas other relations in the original
fermion space, such as e.g.  the trivial operator
equivalence between a bifermion operator and the product of
two single fermion operators, will only hold on the
physical subspace of the ideal space.  (See also
Sec.{\ }\ref{sec3a}.)

A mapping for the full {so(2$N$+1)} would entail the
introduction of a boson (associated operators
$B^{\mu\nu}\equiv B^{\dag}_{\mu\nu}$ and $B_{\mu\nu}$) for
each fermion pair with indices $\mu\nu$, together with
kinematically independent ideal fermions or ideal
quasifermions (associated operators $\alpha^\mu\equiv
\alpha^{\dag}_{\mu}$ and $\alpha_\mu$).  (The distinction
between ideal fermions and ideal quasifermions is linked to
the algebraic structure associated with the corresponding
operators, as elaborated below.)  Kinematic independence
dictates that boson and ideal (quasi)fermion operators
commute,
   \begin{equation}\label{eq:4}
   [B^{\mu\nu},\alpha^{\theta}] = [B^{\mu\nu},\alpha_{\theta}] =0
   \end{equation}
with similar results for the conjugate combinations.
Furthermore, the physical subspace in the {so(2$N$+1)}
case contains states with one ideal fermion at most, since
the bosons $B$ above had been introduced to represent
fermion pair degrees of freedom.

In the physically interesting case where a collective
subalgebra of {so(2$N$+1)} exists, one is really only
interested in bosonizing the corresponding collective
fermion pair(s), while treating all remaining degrees of
freedom as fermions. In this case the ideal space should
therefore not be limited with respect to the number of
ideal fermions.

In the familiar example of pairing in a single $j$-shell
where a single collective boson, $B^\dagger$ (say),
suffices to represent the collective fermion pair, one
would naturally aim at a product space description in terms
of basis states of the type $ (B^{\dagger})^n
\alpha^{\mu_1}{\alpha}^{\mu_2}\ldots|0)$, where the
operators $\alpha$ represent ideal (quasi)fermions.

We recall here that our approach to boson-fermion mappings
resorts under what has broadly been termed \cite{KM91} the
Beliaev-Zelevinsky-Marshalek method in which a mapping of {\it
operators} precedes a mapping of {\it states}.  States are
then mapped after an association of extreme weight states
has been made, usually in the form $|0\rangle\longleftrightarrow|0)$,
as we
also do here. The fermion vacuum $|0\rangle$ is annihilated by
all fermion annihilation operators, $a_\mu|0\rangle=0$, while the
ideal space vacuum is annihilated by all ideal
(quasi)fermion and all boson annihilation operators, namely
$\alpha_\mu|0) = B_{\mu\nu}|0) = 0$.

We now turn to the difference between ideal fermions
and ideal quasifermions, the latter also often referred to
simply as quasifermions \cite{KM91}.  This difference
resides in the way in which single fermion degrees of
freedom in the ideal space take into account information
about the existing or pre-chosen fermion pair -- boson
association \cite{GH80a,KM91}. It is instructive to
illustrate this in the su(2) case where in the ideal space
the single boson degree of freedom $B^\dagger$ represents
the original correlated fermion pair $A^+$. Clearly a
similar configuration of fermions in the ideal space will
be redundant. To take this into account, the algebra of
ideal space fermions may be modified by imposing the {\it
operator} constraint \cite{KM91}
   \begin{equation}
   \sum_{\mu>0}\alpha^\mu \alpha^{\bar \mu} = 0.
   \end{equation}
This results in a modification of the fermion algebra in
the ideal space \cite{KM91}, in which case the
corresponding fermion-like operators are referred to as
(ideal) {\it quasi}-fermion operators.

Alternatively to this procedure it is possible to retain
the usual algebra for the ideal fermions (hence the
corresponding terminology) and to incorporate the
implications of a pre-chosen fermion pair -- boson
association into the ideal space {\it images} of the
original single fermion operators \cite{GH80a,GH83}.

As may be expected intuitively and has been shown
explicitly \cite{GH80a} in the case of mappings for
{so(2$N$+1)}, ideal fermions and quasifermions may be related
on an operator level by showing that the  ideal
quasifermion operators have the form of the corresponding
ideal fermion operator times a projection operator. We
emphasize, however, that a similar relationship has not
yet been identified in detail for
any of the cases where a collective subalgebra dictates the
bosons that appear in the ideal space.

We note here that in the standard phenomenological IBFM it
is indeed ideal fermions (and not quasifermions) that enter
the description. In microscopic analyses which address the
link between phenomenological IBFM parameters and those of
an underlying shell model, present
discrepancies \cite{KC93} between results obtained from a
mapping in terms of ideal fermions \cite{dKG88,GM89} and one
constructed in terms of quasifermions \cite{KC93}, must at
least partially be ascribed to the different algebraic
properties of ideal fermions and quasifermions.

In the sequel we develop our formalism only for ideal
fermions which seem not only to be more naturally suited
for incorporation into coherent states, but also
closer to the spirit in which odd fermions (with unaltered
algebra) are introduced phenomenologically, as discussed
above and in Sec.{\ }\ref{sec1}.

To conclude this Section, we briefly mention an alternative
approach to the same problem, albeit one which mainly
focuses on different or complimentary aspects, namely
vector coherent state theory (VCS) \cite{Hecht,Rowe}.
Although this approach also uses ``intrinsic" degrees of
freedom to account for the odd fermions (ideal
(quasi)fermions above), these degrees of freedom are
utilized much more indirectly than ideal (quasi)fermions
and are only defined in terms of their (left) action on the
vector coherent states, rather than through an explicit
algebraic structure.  Furthermore this approach has so far
mostly been utilized in the context of explicit
construction of matrices for irreducible representations.
It has also proven to be a valuable formalism for
identifying physical subspaces through what is termed
$K$-matrix theory (see Ref.{\ }\cite{Rowe} and references therein).

Aspects of the relationship between the QBV and VCS
approaches have recently been studied by Klein, Walet,
Geyer and Hahne \cite{KWGH91}.

\section{THE {so(2$N$)} BOSON-FERMION MAPPINGS}
\label{sec3a}

The {so(2$N$)} algebra consists of all bifermion operators in a
fermion Fock space built of $N$ single-particle states,
i.e., $a^{\mu}a^{\nu}, a_{\nu}a_{\mu}$, and
$\frac{1}{2}\delta^{\mu}_{\nu}-a^{\mu}a_{\nu}$,
where $a^{\mu}$ and $a_{\mu}$ denote fermion
creation and annihilation operators, respectively,
$a^{\mu}$=$(a_{\mu})^+$.  The {so(2$N$)} superalgebra is obtained by
adding to the {so(2$N$)} algebra the single-fermion operators
themselves.  Their anticommutation relations,
   \begin{equation}\label{e100}
   \{a^{\mu},a_{\nu}\} = \delta_{\nu}^{\mu} ,
   \end{equation}
determine the commutation relations between the
single-fermion and bifermion operators, as well as the
{so(2$N$)} commutation relations between the bifermion
operators.

\subsection{The {so(2$N$)} supercoherent states}
\label{sec3a1}
The {so(2$N$)} supercoherent states can be defined as \cite{SCS}
   \begin{equation}\label{e101}
   |{C,\phi}\rangle =
   \exp\left({\textstyle\frac{1}{2}}
   C_{\mu\nu}a^{\mu}a^{\nu} + \phi_{\mu}a^{\mu}\right) |0\rangle
   \end{equation}
with the usual summation convention applied, and $|0\rangle$
denoting the fermion vacuum.  These supercoherent states
depend on $N(N-1)/2$ complex numbers,
$C_{\mu\nu}$=$-$$C_{\nu\mu}=(C^{\mu\nu})^{\displaystyle\ast}$,
and on $N$ complex
Grassmann variables,
$\{\phi_{\mu},\phi_{\nu}\}$=$\{\phi_{\mu},\phi^{\nu}\}$=0,
$\phi_{\mu}$=$(\phi^{\mu})^{\displaystyle\ast}$, which
{\it anticommute} with the fermion operators,
$\{\phi_{\mu},a_{\nu}\}$=$\{\phi_{\mu},a^{\nu}\}$=0.  The ``bra''
supercoherent state,
   \begin{equation}\label{e102}
   \langle{C,\phi}| =
   \langle{0}| \exp\left({\textstyle\frac{1}{2}}
   C^{\mu\nu} a_{\nu} a_{\mu} + \phi^{\mu} a_{\mu} \right) ,
   \end{equation}
facilitates the construction of a functional representation of
the fermion Fock space. To every many-fermion state
$|\Psi\rangle$ one namely associates a function of variables
$C^{\mu\nu}$ and $\phi^{\mu}$ according to the simple prescription
   \begin{equation}\label{e103}
   |\Psi\rangle \longleftrightarrow f_\Psi(C,\phi) =
   \langle{C,\phi}|\Psi\rangle .
   \end{equation}
Let us now consider the superalgebra composed of $N(N-1)/2$
boson creation and annihilation operators, $B^{\mu\nu}$ and
$B_{\mu\nu}$, $B^{\mu\nu}$=$-$$B^{\nu\mu}=
(B_{\mu\nu})^{\mbox{\footnotesize{\dag}}}$,
and of $N$ ideal
fermion creation and annihilation operators $\alpha^{\mu}$ and
$\alpha_{\mu}$, $\alpha^{\mu}$=$(\alpha_{\mu})^+$, i.e.,
   \begin{equation}\label{e104}\begin{array}{rcl}
   \big[B_{\mu\nu},B^{\theta\rho}\big] &=&
      \delta^{\theta}_{\mu}\delta^{\rho}_{\nu}-
      \delta^{\theta}_{\nu}\delta^{\rho}_{\mu} , \\
   \big[B_{\mu\nu},\alpha^{\mu}\big] &=&
   \big[B_{\mu\nu},\alpha_{\mu}\big] = 0 , \\
   \left\{\alpha^{\mu},\alpha_{\nu}\right\} &=& \delta^{\mu}_{\nu},\\
   \left\{\alpha^{\mu},\alpha^{\nu}\right\} &=& 0                . \\
   \end{array}\end{equation}
We refer to $\alpha^{\mu}$ as ideal fermions to distinguish
them from real fermions $a^{\mu}$. The appellation
``ideal'' serves
as a reminder that the creation operators $\alpha^{\mu}$ {\em
commute} with the boson annihilation operators $B_{\mu\nu}$,
cf.{\ }Eq.{\ }(\ref{eq:4}),
as opposed to the real fermion creation operators $a^{\mu}$
which {\em do not} commute with pair annihilation operators
$a_{\mu} a_{\nu}$.

The supercoherent state for the superalgebra (\ref{e104}),
   \begin{equation}\label{e105}
   |{C,\phi}) =
   \exp\left({\textstyle\frac{1}{2}}
   C_{\mu\nu} B^{\mu\nu} + \phi_{\mu} \alpha^{\mu} \right) |0) ,
   \end{equation}
where $|0)$ denotes the ideal boson-fermion vacuum,
$B_{\mu\nu}|0)$=$\alpha_{\mu}|0)$=0, gives rise to a functional
representation of the ideal boson-fermion states:
   \begin{equation}\label{e106}
   |\Psi) \longleftrightarrow f_\Psi(C,\phi) = ({C,\phi}|\Psi) .
   \end{equation}
We apply the usual notation by denoting the real fermion
states and the ideal boson-fermion states by angled and
rounded brackets, respectively.  By comparing
Eqs.{\ }(\ref{e103}) and (\ref{e106}) we see that both real
and ideal states are now represented as functions of
variables $C^{\mu\nu}$ and $\phi^{\mu}$, which provides us with a
powerful method of mapping real fermion states into the
ideal boson-fermion states (cf.{\ }Ref.{\ }\cite{Dob81a}).
Indeed, we may define the
boson-fermion image of a fermion state by requiring that
their functional images are equal, i.e.,
   \begin{equation}\label{e200}
   |\Psi) \longleftrightarrow |\Psi\rangle
   \end{equation}
if
   \begin{equation}\label{e201}
   ({C,\phi}|\Psi) \equiv \langle{C,\phi}|\Psi\rangle .
   \end{equation}

\subsection{The {so(2$N$)} Usui operator}
\label{sec3a2}
All subsequent constructions of mappings between fermion
operators and functions of boson and ideal fermion operators
can be carried out as indicated above.
One can, however, avoid the functional representation
as an
intermediate step in the mapping procedure by alternatively
considering
the supercoherent-state-inspired generalized Usui operator
(see also Refs.{\ }\cite{Usui,DJ73})
   \begin{equation}\label{e107}
   U = \langle{0}|\exp\left({\textstyle\frac{1}{2}}
       B^{\mu\nu} a_{\nu} a_{\mu} +
             \alpha^{\mu} a_{\mu}\right)|0) .
   \end{equation}
This operator transforms a real fermion state into an ideal
boson-fermion state
   \begin{equation}
   |\Psi) = U |\Psi\rangle
   \end{equation}
in such a way that, Eq.{\ }(\ref{e201}) holds automatically.
Note that in defining the generalized Usui operator as
in Eq.{\ }(\ref{e107}) we imply that the ideal fermion operators
$\alpha^{\mu}$ and $\alpha_{\mu}$ {\it anticommute} with the real
fermion operators $a^{\nu}$ and $a_{\nu}$.  By using the Usui
operator one effectively avoids dealing with Grassmann variables
which have rather unconventional properties, especially when one
concerns derivatives with respect to Grassmann
variables. However, reference to the supercoherent state
(\ref{e105}) and the functional images remains useful, as also
becomes clear from the subsequent discussion. (In Appendix \ref{appC}
we also give an explicit example of how functional images
are utilized to derive operator mappings.)

The mapping between operators acting in the real and ideal
spaces can thus be realized by exploiting the Usui operator
(\ref{e107}). If for a real
fermion operator $\hat O$ one can find an operator ${\cal O}$
acting in the ideal space such that
   \begin{equation}\label{e108}
   {\cal O} U = U\hat O ,
   \end{equation}
we say that $\hat O$ is mapped to ${\cal O}$, i.e.
${\cal O}$ is the boson-fermion image of
$\hat O$ under the mapping:
   \begin{equation}\label{e109}
   {\cal O}  \longleftrightarrow \hat O .
   \end{equation}
Such a definition does not determine properties of ${\cal O}$ in
the full ideal space, but only those pertinent to the
so-called physical subspace which consists of images
$U|\Psi\rangle$ of all real fermion states
$|\Psi\rangle$. Therefore, in the full ideal space the
boson-fermion image of a fermion
operator is not unique.

In Appendix \ref{appA} we derive the following
boson-fermion mapping of fermion and bifermion operators as
determined by the Usui operator of Eq.{\ }(\ref{e107}):
   \begin{mathletters}\begin{eqnarray}
   a^{\mu} a^{\nu} &\longleftrightarrow&
                       B^{\mu\nu}
                 - B^{\mu\rho} B^{\nu\theta} B_{\rho\theta}
                                \nonumber     \\
              &    & - B^{\mu\rho}\alpha^{\nu}\alpha_{\rho}
                     + B^{\nu\rho}\alpha^{\mu}\alpha_{\rho}
                     + \alpha^{\mu}\alpha^{\nu} ,
                                \label{e110c} \\
   a^{\mu} a_{\nu} &\longleftrightarrow&
                       B^{\mu\theta} B_{\nu\theta}
                 + \alpha^{\mu}\alpha_{\nu} ,
                                \label{e110b} \\
   a_{\nu} a_{\mu} &\longleftrightarrow&
                       B_{\mu\nu}               ,
                                \label{e110a} \\
   a^{\nu}         &\longleftrightarrow&
                       \alpha^{\nu}
                 + B^{\nu\rho}\alpha_{\rho} ,
                                \label{e110e} \\
   a_{\nu}      &\longleftrightarrow&
                       \alpha_{\nu}             .
                                \label{e110d}
   \end{eqnarray}\end{mathletters}
It should be stressed that once the Usui operator is defined,
the mapping of operators is also uniquely defined through
Eq.{\ }(\ref{e108}), and the mappings (\ref{e110c})--(\ref{e110d})
result from a simple calculation.

The images of superalgebra generators, obtained by using the
Usui operator (\ref{e107}), are by construction guaranteed
to fulfil the (anti)commutation relations only in the
physical space. However, in the functional representation,
these images have a particularly simple form containing only
first order
differential operators. In the ideal space, this means that
only a single boson (or a single fermion)
annihilation operator appears in
any of the images in Eqs.{\ }(\ref{e110c})--(\ref{e110d}).
Together with the fact that the single-boson and
single-ideal-fermion states are indeed physical, this
ensures that the mapped operators fulfil (anti)commutation
relations in the entire ideal space
(cf.{\ }discussion in Sec.{\ }2
of Ref.{\ }\cite{Dob81b}).
Of course, this fact can also be checked {\em a posteriori}
by explicitly verifying the {so(2$N$)} superalgebra
(anti)commutation relations of the operators in
Eqs.{\ }(\ref{e110c})--(\ref{e110d}).

The latter fact ensures that the boson-fermion images
of real fermion states do not depend on the way
we group fermion operators before we
construct ideal states by consecutively acting with operator
images in the ideal space. For example, one may obtain
the boson-fermion image of the state
$a^{\mu} a^{\nu}|0\rangle$ either by acting with the image
of $a^{\mu} a^{\nu}$, Eq.{\ }(\ref{e110c}), on the boson-fermion
vacuum $|0)$, or by acting twice with images
of single-fermion operators, Eq.{\ }(\ref{e110e}).
The final result is the same in both cases, and a similar
conclusion also holds in more complicated cases.

On the other hand, there is no guarantee that the
image of a product of real fermion operators is
equal to the product of their images. In general,
this equality does not hold in the operator sense,
but of course it does when action on a physical
state is considered.

One notes the appearance of the ideal fermion pair
$\alpha^{\mu}\alpha^{\nu}$ in the mapping of the real fermion
pair $a^{\mu} a^{\nu}$, Eq.{\ }(\ref{e110c}).
Therefore, the zero-, one-, and two-fermion
states have the following ideal boson-fermion images:
   \begin{mathletters}\begin{eqnarray}
                  |0\rangle &\longleftrightarrow&
                                  |0) , \label{e202a} \\
           a^{\nu}|0\rangle &\longleftrightarrow& \alpha^{\nu}
                                  |0) , \label{e202b} \\
   a^{\mu} a^{\nu}|0\rangle &\longleftrightarrow& \left[B^{\mu\nu}
                                        + \alpha^{\mu}\alpha^{\nu}
                                                  \right]
                            |0) . \label{e202c}
   \end{eqnarray}\end{mathletters}
The real fermion pairs are thus mapped onto linear combinations
of ideal bosons and ideal fermion pairs. The mapping
faithfully reproduces the structure of the real fermion space,
i.e., only the symmetric combinations,
$B^{\mu\nu} + \alpha^{\mu}\alpha^{\nu}$,
appear in the physical space, while
the antisymmetric ones,
$B^{\mu\nu} - \alpha^{\mu}\alpha^{\nu}$,
belong to the unphysical space.

As discussed in Sec.{\ }\ref{sec1}, the mapping of fermion
states onto the ideal boson-fermion space aims at such a
description of Pauli correlations between even core and an
odd particle which avoids explicit antisymmetrization. From
this point of view, the mapping in
Eqs.{\ }(\ref{e110c})--(\ref{e110d}) does not represent any
gain with respect to the original fermion space.  Images of
even fermion states, obtained by acting on the vacuum with
the images of $a^{\mu} a^{\nu}$, Eq.{\ }(\ref{e110c}), contain the
ideal fermion pair $\alpha^{\mu}\alpha^{\nu}$,
cf.{\ }Eq.{\ }(\ref{e202c}),
and an explicit
antisymmetrization with any odd ideal fermion
is still required.
This is not a satisfactory solution, because
one would like to achieve a complete bosonization
of the real even-fermion-number states,
similarly as is the case for the usual Dyson mapping, where
ideal fermions are not used.
In the following sections we discuss methods of addressing
this deficiency.

\subsection{Similarity-transformed {so(2$N$)} \\
              boson-fermion mappings}
\label{sec3a4}
By applying a similarity transformation ${\cal W}$
to all images of superalgebra generators,
${\cal O}'={\cal W}^{-1}{\cal O}{\cal W}$,
one obtains another possible
mapping of the superalgebra in the ideal space.
This corresponds to using a new Usui operator,
$U'={\cal W}U$, and the new
physical space is then equal to the similarity transform
of the original physical space,
$|\Psi)'=U'|\Psi\rangle={\cal W}|\Psi)$.
A suitable choice of the similarity
transformation may therefore change the composition
and properties of the physical space, and  lead to
mappings with a structure closer to the structures envisaged
in Sec.{\ }\ref{sec1}. In what follows we particularly aim
at removing the unwelcome term $\alpha^{\mu}\alpha^{\nu}$
through an appropriate similarity transformation.

The similarity transformation $\cal W$ can always be
presented in the form of an exponent,
${\cal W}=e^{{\cal T}}$, and evaluated by
applying the Baker-Campbell-Hausdorf formula,
   \begin{eqnarray}
   {\cal W}^{-1}{\cal O}{\cal W} &=&
               e^{-{\cal T}} {\cal O} e^{{\cal T}} \nonumber \\
                   &=& \sum_{k=0}^\infty \frac{(-1)^k}{k!}
                       [{\cal T}[{\cal T}\ldots[{\cal T},{\cal O}]
                                     \ldots]]_k , \label{e114}
   \end{eqnarray}
where the multiple commutator has to be taken $k$ times.
The series is infinite unless the multiple
commutator vanishes at some order. Since we would like to
preserve
the finiteness of the boson mapping, we will consider only
such operators ${\cal T}$ which lead to finite series in
Eq.{\ }(\ref{e114}).
Below we
present results for two specific operators ${\cal T}$, while
some details of the derivation are given in Appendix
\ref{appB}.

Let us first discuss the similarity transformation
(\ref{e114}) with ${\cal T}$ given by
   \begin{equation}\label{e115}
   {\cal T} = {\textstyle\frac{1}{2}}
      B^{\mu\nu}\alpha_{\nu}\alpha_{\mu} ,
   \end{equation}
which, when applied to mapping (\ref{e110c})--(\ref{e110d}),
yields
   \begin{mathletters}\begin{eqnarray}
   a^{\mu} a^{\nu}  &\longleftrightarrow&
                           \alpha^{\mu}\alpha^{\nu}
                   - B^{\mu\rho} B^{\nu\theta} B_{\rho\theta} ,
                                \label{e116c} \\
   a^{\mu} a_{\nu}  &\longleftrightarrow&
                           \alpha^{\mu}\alpha_{\nu}
                   + B^{\mu\theta} B_{\nu\theta}             ,
                                \label{e116b} \\
   a_{\nu} a_{\mu}  &\longleftrightarrow&
                           \alpha_{\nu}\alpha_{\mu} + B_{\mu\nu}   ,
                                \label{e116a} \\
   a^{\nu}          &\longleftrightarrow& \alpha^{\nu}             ,
                                \label{e116e} \\
   a_{\nu}          &\longleftrightarrow& \alpha_{\nu}             .
                                \label{e116d}
   \end{eqnarray}\end{mathletters}
One can see that the effect obtained is exactly the
opposite to the desired one. Namely, the boson-fermion
image of the real fermion pair operator $a^{\mu} a^{\nu}$,
Eq.{\ }(\ref{e116c}), creates solely the ideal fermion pairs
$\alpha^{\mu}\alpha^{\nu}$ when acting on the ideal vacuum, and
the bosons do not at all appear in the physical space. The
mapping in Eqs.{\ }(\ref{e116c})--(\ref{e116d}) simply
replaces real fermions by the ideal ones, and is therefore
useless for practical applications.

On the other hand, mapping (\ref{e116c})--(\ref{e116d})
may serve for an explicit check of
some properties of other similarity transformed images.
For example, it trivially fulfils the {so(2$N$)} superalgebra
(anti)commutation relations in the whole ideal space.
Also trivially, the image of
any many-fermion state $a^{\mu} a^{\nu}\ldots a^{\rho}|0\rangle$
is always $\alpha^{\mu}\alpha^{\nu}\ldots\alpha^{\rho}|0)$,
no matter in which way we group (or do not group) the fermion
operators in pairs to use either the image of
$a^{\mu} a^{\nu}$, Eq.{\ }(\ref{e116c}), or that of $a^{\nu}$,
Eq.{\ }(\ref{e116e}).
Therefore, any similarity transformed mapping will also
have these properties.

{}From the above result one can guess that the desired
goal may be met by using the hermitian conjugate of the
operator in Eq.{\ }(\ref{e115}) to construct the similarity
transformation.  In Appendix \ref{appB} we show that by
transforming the mapping (\ref{e110c})--(\ref{e110d}) with
   \begin{equation}\label{e117}
   {\cal T}={\cal T}({\cal X}) \quad\mbox{for}\quad
   {\cal X}={\textstyle\frac{1}{2}}\alpha^{\mu}\alpha^{\nu} B_{\mu\nu}
   \end{equation}
one obtains
   \FL\begin{mathletters}\begin{eqnarray}
   a^{\mu} a^{\nu} &\longleftrightarrow&
                       B^{\mu\nu}
                 - B^{\mu\rho} B^{\nu\theta} B_{\rho\theta}
                                \nonumber     \\
              &    & - B^{\mu\rho}\alpha^{\nu}\alpha_{\rho}
                     + B^{\nu\rho}\alpha^{\mu}\alpha_{\rho}
                     - 2\alpha^{\mu}\alpha^{\nu}{\cal T}'{\cal N} ,
                                \label{e119c} \\
   a^{\mu} a_{\nu} &\longleftrightarrow&
                       B^{\mu\theta} B_{\nu\theta}
                 + \alpha^{\mu}\alpha_{\nu}                   ,
                                \label{e119b} \\
   a_{\nu} a_{\mu} &\longleftrightarrow&
                      B_{\mu\nu}                                  ,
                            \label{e119a} \\
   a^{\nu}         &\longleftrightarrow&
                      \alpha^{\nu}{\cal T}'(1-{\cal N})
                - \alpha^{\rho}{\cal T}' B^{\nu\theta}
                                         B_{\rho\theta}
                    + B^{\nu\rho}\alpha_{\rho}                   ,
                                \label{e119e} \\
   a_{\nu}         &\longleftrightarrow&
                      \alpha_{\nu}
                + \alpha^{\rho} B_{\nu\rho}{\cal T}'         .
                                \label{e119d}
   \end{eqnarray}\end{mathletters}
In these equations, ${\cal N}$ is the ideal fermion number
operator,
   \begin{equation}\label{e118}
   {\cal N} = \alpha^{\mu}\alpha_{\mu} ,
   \end{equation}
while ${\cal T}$ is an analytical function of ${\cal X}$,
   \begin{equation}\label{e120}
   {\cal T} = \sum_{k=0}^\infty \lambda_k{{\cal X}}^k ,
   \end{equation}
whose first derivative ${\cal T}'$ obeys the Ricatti equation
\cite{Ricatti},
   \begin{equation}\label{e121}
   2{\cal X}({\cal T}''+{{\cal T}'}^2) + {\cal T}' - 1 = 0 .
   \end{equation}
This particular Ricatti equation can be solved in a closed
form, and one obtains
   \begin{equation}\label{e122}\begin{array}{rcl}
   {\cal T}'&=&\left(\tanh\sqrt{2{\cal X}}\right)/\sqrt{2{\cal X}},\\
   {\cal T} &=&\log\cosh\sqrt{2{\cal X}} , \\
   \end{array}\end{equation}
which gives the similarity transformation
   \begin{equation}\label{e123}
   {\cal W} = \cosh\sqrt{\alpha^{\mu}\alpha^{\nu} B_{\mu\nu}} .
   \end{equation}
The square roots of operators, which appear in the above
expressions, only serve as a short-hand notation to
describe the power series.  In fact, all these series
contain only even powers of the argument, and therefore are
the series of powers of the
${\cal X}$=$\frac{1}{2}\alpha^{\mu}\alpha^{\nu}B_{\mu\nu}$
operator itself (without the square root).  For
example, the lowest-order terms of the operator ${\cal T}'$,
which enters mapping (\ref{e119c})--(\ref{e119d}), read
   \begin{equation}\label{e204}
   {\cal T}'=1-{\textstyle\frac{1} {3}}
                    \alpha^{\mu}\alpha^{\nu} B_{\mu\nu}
              +{\textstyle\frac{2}{15}}
                  \left(\alpha^{\mu}\alpha^{\nu} B_{\mu\nu}
                       \right)^2 - \ldots
   \end{equation}

The functions of ${\cal X}$ are in principle infinite power
series.  However, convergence problems for these functions
never appear if
one considers their action on ideal states with a given
number of bosons.  Indeed, an $n$-th power of ${\cal X}$
annihilates all boson states which have a boson number
smaller then $n$.  Therefore, the infinite power series can
be cut off at the $n$-th term, whenever only such ideal
states are considered.

{}From Eq.{\ }(\ref{e119c}) one sees that the
even-fermion-number states are now entirely bosonized.
This is so because the last three terms in this equation
give a contribution only if an ideal fermion is already
present, while they can be disregarded in a pure boson
subspace.  Therefore, the images of the even fermion states
reduce to those given by the standard Dyson mapping
\cite{JDF71}, for which the
mapping of the one- to four-fermion states has now the
following explicit form:
   \FL\begin{mathletters}\begin{eqnarray}
                                       |0\rangle &\longleftrightarrow&
                                               |0) , \label{e203a} \\
                                a^{\nu}|0\rangle &\longleftrightarrow&
                             \alpha^{\nu}|0) , \label{e203b} \\
                        a^{\mu} a^{\nu}|0\rangle &\longleftrightarrow&
                               B^{\mu\nu}  |0) , \label{e203c} \\
             a^{\lambda}a^{\mu} a^{\nu}|0\rangle &\longleftrightarrow&
                                   \left[
                            \alpha^{\lambda}B^{\mu\nu}
                                   -\alpha^{\mu}    B^{\lambda\nu}
                                   +\alpha^{\nu}    B^{\lambda\mu}
                                        \right]|0)   \nonumber     \\
                                            &    &
                                   -2\alpha^{\lambda}
                                    \alpha^{\mu}\alpha^{\nu}
                                               |0) , \label{e203d} \\
   a^{\kappa}a^{\lambda}a^{\mu} a^{\nu}|0\rangle &\longleftrightarrow&
                                   \left[
                                    B^{\kappa\lambda}B^{\mu\nu}
                                   -B^{\kappa\mu}    B^{\lambda\nu}
                                        \right.       \nonumber     \\
                                            &    & \left.
                          \phantom{ B^{\kappa\lambda}B^{\mu\nu} +}
                                   +B^{\kappa\nu}    B^{\lambda\mu}
                                        \right]|0) . \label{e203e}
   \end{eqnarray}\end{mathletters}

When an odd fermion is added to an even fermion state, the
last term in Eq.{\ }(\ref{e119e}) does not contribute and
the first two terms create an odd ideal fermion.  However,
this odd ideal fermion is accompanied by a whole series of
terms created by the operator ${\cal T}'$.  Therefore, the image
of an odd real fermion state is a mixture of one-, three-,
five-, e.t.c.  ideal fermion states.  More precisely, the
series continues till the number of ideal fermions reaches
the number of real fermions in the odd state being mapped.

This is exemplified in the image of the three-fermion
state, Eq.{\ }(\ref{e203d}), which contains a
three-ideal-fermion component.  On the other hand, the
one-ideal-fermion component of this image is built as an
antisymmetrized product of the ideal fermion and of the boson
representing the even core.  The structure of odd states
with more particles is similar.

When a second odd fermion is added to an odd state, all
ideal fermions disappear by automatically
{\it recombining} to bosons.  This is not at all evident
when looking at the  rather involved structure of the
single-fermion image, Eq.{\ }(\ref{e119e}), which contains
an infinite series of terms {\em creating} ideal
fermions.  However, the odd state is itself built as a
series of terms with different ideal-fermion numbers.  Both
series conspire in such a way that the recombination
mechanism is perfectly realized and the Pauli correlations
exactly preserved.

The operator ${\cal T}'$ can thus be regarded as an operator
responsible for the necessary antisymmetrization between
ideal fermions and bosons.  An approximate
antisymmetrization can be achieved by neglecting higher
order terms in the series expansion (\ref{e204}).  When
keeping terms up to the $n$-th order one assures a correct
antisymmetrization of states with the number of bosons not
greater then $n$.

The exact preservation of Pauli correlations have
been achieved here at the expense of complicated images
of operators. When the mapped operators
(\ref{e119c})--(\ref{e119d}) are e.g. applied to
a (real) fermion Hamiltonian, one obtains its
boson-fermion image which acquires
many-body terms. One may then separate terms
into the boson-boson, fermion-fermion, and
boson-fermion parts and therefore split the
Hamiltonian into three parts, as in Eq.{\ }(\ref{e2}).
The boson-fermion part then represents an
interaction which enforces correct antisymmetrization
between the even core and odd fermions.

The exact images of odd states, obtained in this section,
are probably too complicated to be effectively used in
practical calculations. Our ultimate goal which, in the
context of the full {so(2$N$+1)} algebra,
was to describe odd states in a product space of
a {\em single} ideal fermion and bosons, has not yet been met.
On the other hand, we may split the boson-fermion images of odd
states into components having different ideal-fermion
numbers, and consider them separately.
Since these components are all orthogonal one to another,
the antisymmetry properties must be valid for every
one of them. In this way we may consider the images
of odd states {\em projected} on the single-ideal-fermion
subspace as the result of the mapping. Such a projection
is not, of course, a similarity transformation and some
properties of the mapping may therefore be modified.
We analyze these questions in the next section.

\subsection{Projected {so(2$N$)} boson-fermion mapping}
\label{sec3a3}
Apart from the term $\alpha^{\mu}\alpha^{\nu}$ in
Eq.{\ }(\ref{e110c}), the mapping of bifermion
operators, Eqs.{\ }(\ref{e110c})--(\ref{e110a}), is
identical to that derived by D\"onau and Janssen
\cite{DJ73}. They have used the Usui operator which is a
projection of that of Eq.{\ }(\ref{e107}) on the ideal space
with at most one ideal fermion, i.e.,
   \begin{eqnarray}
   U_{01} &\equiv& {\cal P}_{01} U \nonumber \\
          &  =   & \langle{0}|\exp\left({\textstyle\frac{1}{2}}
                            B^{\mu\nu} a_{\nu} a_{\mu}\right)
                            \left(1 + \alpha^{\mu} a_{\mu}\right)|0) ,
                         \label{e111}
   \end{eqnarray}
where ${\cal P}_{01}={\cal P}_0 + {\cal P}_1$, and
${\cal P}_0$=$|0)(0|$ and
${\cal P}_1$=$\alpha^{\mu}|0)(0|\alpha_{\mu}$
are projection operators on the vacuum and on the one-fermion
ideal states, respectively.
Such an Usui operator maps real fermion operators
according to the prescription
   \begin{equation}\label{e112}
   {\cal O}_{01} U_{01} = U_{01}\hat O ,
   \end{equation}
where the image of $\hat O$ is denoted by ${\cal O}_{01}$.
Hence, the mapping of the {so(2$N$)} superalgebra reads
   \begin{mathletters}\begin{eqnarray}
   a^{\mu} a^{\nu}  &\longleftrightarrow&
                        B^{\mu\nu}
                  - B^{\mu\rho} B^{\nu\theta} B_{\rho\theta}
                                                   \nonumber     \\
               &    & - B^{\mu\rho}\alpha^{\nu}\alpha_{\rho}
                      + B^{\nu\rho}\alpha^{\mu}\alpha_{\rho}    ,
                                                   \label{e113c} \\
   a^{\mu} a_{\nu}  &\longleftrightarrow&
                        B^{\mu\theta} B_{\nu\theta}
                  + \alpha^{\mu}\alpha_{\nu}                ,
                                                   \label{e113b} \\
   a_{\nu} a_{\mu}  &\longleftrightarrow&
                        B_{\mu\nu}                              ,
                                                   \label{e113a} \\
   a^{\nu}          &\longleftrightarrow&
                        \left(\alpha^{\nu}
                  - \alpha^{\rho} B^{\nu\theta}
                     B_{\rho\theta}\right){\cal Q}
                       + B^{\nu\rho}\alpha_{\rho}               ,
                                                   \label{e113e} \\
   a_{\nu}          &\longleftrightarrow&
                        \alpha_{\nu}
                  + \alpha^{\rho} B_{\nu\rho}{\cal Q}       .
                                                   \label{e113d}
   \end{eqnarray}\end{mathletters}
Here ${\cal Q}$ denotes an arbitrary operator which
conserves the vacuum and
annihilates one-ideal-fermion states, i.e.,
   \begin{equation}\label{e205}
   {\cal Q} = {\cal P}_0 +{\cal Q}'(1-{\cal P}_0)(1-{\cal P}_1)  ,
   \end{equation}
where ${\cal Q}'$ is arbitrary.

The images of the {so(2$N$)} generators,
Eqs.{\ }(\ref{e113c})--(\ref{e113d}) can be derived
in two ways. First,
one may follow a direct and standard way
(see Appendix \ref{appC})
of explicitly considering the projected Usui operator,
Eq.{\ }(\ref{e111}). Second, one may perform
a kind of projection of the similarity images ${\cal O}$,
Eqs.{\ }(\ref{e119c})--(\ref{e119d}),
by using the equation
   \begin{equation}\label{e206}
   {\cal P}_{01}{\cal O} = {\cal O}_{01}{\cal P}_{01} ,
   \end{equation}
to find ${\cal O}_{01}$.

Equation (\ref{e206}) results from
the definitions of boson-fermion
images, Eqs.{\ }(\ref{e108}) and (\ref{e112}),
and the relationship between the corresponding
Usui operators (\ref{e111}). In particular, it has
the following solutions for ideal fermions
in the similarity mapping:
   \begin{equation}\begin{array}{rll@{}rl}\label{e207}
   &{\cal P}_{01}\alpha^{\nu}
       &= & \alpha^{\nu}       {\cal Q}{\cal P}_{01}& , \\
   &{\cal P}_{01}\alpha_{\nu}
       &= & \alpha_{\nu}         {\cal P}_{01}& , \\
   &{\cal P}_{01}\alpha^{\mu}\alpha_{\nu}
       &= & \alpha^{\mu}\alpha_{\nu}{\cal P}_{01}& , \\
   &{\cal P}_{01}{\cal N}
       &= &                 {\cal N}{\cal P}_{01}& , \\
   &{\cal P}_{01}{\cal T}'
       &= &                    {\cal P}_{01}& , \\
   &{\cal P}_{01}\alpha^{\mu}\alpha^{\nu}
       &=0&                            & . \\
   \end{array}\end{equation}

The mapping of the single-fermion operators,
Eqs.{\ }(\ref{e113e}) and (\ref{e113d}), is the same as
obtained by Geyer and Hahne \cite{GH80a}, who have used
for ${\cal Q}$ simply the vacuum projection operator,
${\cal Q}$=${\cal P}_0$. Another possible choice is
${\cal Q}$=$1-{\cal N}$, where ${\cal N}$
is the ideal-fermion-number operator,
Eq.{\ }(\ref{e118}).
That ${\cal Q}$ is not unique, simply illustrates the fact that
images of fermion operators in the ideal space are
undetermined outside the physical space, which here
consists only of zero- and one-ideal-fermion states. By the
same token, the superalgebra (anti)commutation relations of
the generator images in the ideal space,
Eqs.{\ }(\ref{e113c})--(\ref{e113d}), are fulfilled only in
the physical space.

The mapping given in
Eqs.{\ }(\ref{e113c})--(\ref{e113d}) presents a satisfactory
solution to the bosonization program presented
in Sec.{\ }\ref{sec1}.
Starting from the vacuum $|0)$ the even
fermion states are obtained by using the image of $a^{\mu}
a^{\nu}$, Eq.{\ }(\ref{e113c}), and therefore are mapped on
purely bosonic states. Then, the odd fermion is simply
added on top of the bosonic state by using the image of
single-fermion creation operator, Eq.{\ }(\ref{e113e}).

On the other hand, when an additional fermion is added
to an odd-fermion state by acting again with the image
of Eq.{\ }(\ref{e113e}), the presence of the
projection operator ${\cal Q}$ assures that the odd fermion
is annihilated and a boson created. This is a concrete
realization of the recombination mechanism described in the previous
section.

\section{BOSON-FERMION MAPPING \\
          OF COLLECTIVE SPACE}
\label{sec3b}

In this section we concentrate our discussion on the
collective subalgebra based on using the collective
fermion-pair creation operators
   \begin{equation}\label{e4}
   A^i = {\textstyle\frac{1}{2}}\chi^i_{\mu\nu} a^{\mu} a^{\nu} ,
   \end{equation}
numbered by the collective index $i$=$1,\ldots,M$, where
$M$ is supposed to be much smaller than the number of all
possible pairs ($N(N-1)/2$).  Together with the
corresponding collective fermion-pair annihilation
operators, $A_i$=$(A^i)^+$, all linearly independent
commutators $\left[A_i,A^j\right]$, and the single-fermion
operators, $a^{\nu}$ and $a_{\nu}$, they are assumed to form a
closed collective superalgebra. The closure conditions read
   \begin{equation}\begin{array}{rcl}\label{e5x}
   \left[\left[A_i,A^j\right],A_k\right] &=& c^{jl}_{ik}A_l
                                           , \\
   \left[A^i,a_{\nu}\right] &=& \chi^i_{\mu\nu} a^{\mu}
                                           , \\
   \left[A^i,a^{\nu} \right] &=& 0
                                           , \\
   \left\{a^{\mu},a_{\nu}\right\} &=& \delta^{\mu}_{\nu}
                                           , \\
   \left\{a^{\mu},a^{\nu} \right\} &=& 0
                                           , \\
   \end{array}\end{equation}
where $c^{jl}_{ik}$ are structure constants and the
implicit summation over repeated collective index $l$ is
assumed.

The corresponding physical subspace of the ideal space is
now envisaged to be
comprised of ideal states with an arbitrary number of
ideal fermions, of course still subject to reigning
space limitations. Physically this reflects a description
where only collective fermion pairs are bosonized, while
all other fermion degrees of freedom are simply
accommodated as ideal fermions.

{}Following Ref.{\ }\cite{DGH91}, we assume that the
collective pairs are orthogonal and normalized to a common
number $g$, i.e.,
   \begin{equation}\label{e60}
   \langle 0| A_i A^j|0\rangle \equiv
        {\textstyle\frac{1}{2}}\chi_i^{\mu\nu}\chi^j_{\mu\nu}
        = g\delta^j_i ,
   \end{equation}
$\big(\chi_i^{\mu\nu}$=$(\chi^i_{\mu\nu})^{\displaystyle\ast}\big)$
which gives the commutation relation
   \begin{equation}\label{e61}
   \left[A^j,\left[A_i,A^l\right]\right] = c^{jl}_{ik}A^k
   \end{equation}
and the symmetry properties of structure constants
   \begin{equation}\label{e62}
   c^{jl}_{ik} = c^{lj}_{ik} =  c^{jl}_{ki}
                             = (c^{ik}_{jl})^{\displaystyle\ast} .
   \end{equation}

\subsection{Collective supercoherent states}
\label{sec3b5}
The collective supercoherent state is defined as
   \begin{equation}\label{e124}
   |{C,\phi}\rangle =
      \exp\left(C_i A^i + \phi_{\mu} a^{\mu}\right)|0\rangle
   \end{equation}
where $C_i$=$(C^i)^{\displaystyle\ast}$ are complex numbers and
$\phi_{\mu}$=$(\phi^{\mu})^{\displaystyle\ast}$
complex Grassmann variables,
as described in Sec.{\ }\ref{sec3a1}. This supercoherent
state suggests the collective Usui operator
   \begin{equation}\label{e125}
   U = \langle{0}|\exp\left(B^i A_i + \alpha^{\mu} a_{\mu}\right)|0)
   \end{equation}
which transforms collective even-fermion states, {\it and}
collective states with added individual fermions, into an
ideal space composed of collective bosons,
$B^i$=$B_i^{\mbox{\footnotesize{\dag}}}$,
$[B^i,B_j]$=$\delta^i_j$, and of ideal
fermions $\alpha^{\mu}$.

The mapping of operators can be obtained from the equation
${\cal O} U$=$U\hat O$, which gives the following mapping of the
collective superalgebra, (\ref{e5x}):
   \begin{mathletters}\begin{eqnarray}
   A^j               &\longleftrightarrow&
                         gB^j - {\textstyle\frac{1}{2}}
                     c^{jl}_{ik}B^i B^k B_l
                                                \nonumber     \\
                &    & - \chi^j_{\mu\rho}\chi_i^{\nu\rho}
                     B^i\alpha^{\mu}\alpha_{\nu}
                              + {\textstyle\frac{1}{2}}
                   \chi^j_{\mu\nu}\alpha^{\mu}\alpha^{\nu}
                                              , \label{e126c} \\
   \big[A_i,A^j\big] &\longleftrightarrow&
                         g\delta^j_i - c^{jl}_{ik}B^kB_l
                       - \chi^j_{\mu\rho}\chi_i^{\nu\rho}
                     \alpha^{\mu}\alpha_{\nu}
                                              , \label{e126b} \\
   A_j               &\longleftrightarrow&
                         B_j                  , \label{e126a} \\
   a^{\nu}           &\longleftrightarrow&
                         \alpha^{\nu}
                   + \chi_i^{\nu\rho} B^i\alpha_{\rho}
                                              , \label{e126e} \\
   a_{\nu}           &\longleftrightarrow&
                         \alpha_{\nu}
                                              . \label{e126d}
   \end{eqnarray}\end{mathletters}
Defining the collective pairs of ideal fermions,
   \begin{equation}\label{e127}
   {\cal A}^j = {\textstyle\frac{1}{2}}
        \chi^j_{\mu\nu}\alpha^{\mu}\alpha^{\nu} ,
   \end{equation}
${\cal A}_j$=$({\cal A}^j)^{\mbox{\footnotesize{\dag}}}$,
one can present the above mapping in a
form in which the pair amplitudes $\chi^j_{\mu\nu}$ do not
appear explicitly:
   \FL\begin{mathletters}\begin{eqnarray}
   A^j               &\longleftrightarrow&
                            {\cal A}^j - {\textstyle\frac{1}{2}}
                      c^{jl}_{ik}B^i B^k B_l
                          + B^i\big[{\cal A}_i,{\cal A}^j\big]
                                              , \label{e128c} \\
   \big[A_i,A^j\big] &\longleftrightarrow&
                            \big[{\cal A}_i,{\cal A}^j\big]
                    - c^{jl}_{ik}B^kB_l
                                              , \label{e128b} \\
   A_j               &\longleftrightarrow&
                            B_j               , \label{e128a} \\
   a^{\nu}           &\longleftrightarrow&
                            \alpha^{\nu}
                    + B^i\big[{\cal A}_i,\alpha^{\nu}\big]
                                              ,  \label{e128e} \\
   a_{\nu}           &\longleftrightarrow&
                            \alpha_{\nu}      .  \label{e128d}
   \end{eqnarray}\end{mathletters}

Similarly as in the {so(2$N$)} case, the image of the
collective pair operator $A^i$,
Eq.{\ }(\ref{e128c}), contains the corresponding
ideal collective pair operator ${\cal A}^i$, and therefore
the above mapping does not present any simplification
in the description of Pauli correlations.
In particular, the collective one-pair states are
not bosonized,
$A^i|0\rangle=({\cal A}^i+B^i)|0)$.
In the following Section we again use a similarity
transformation to remove the intruding term ${\cal A}^i$ from
the image of $A^i$.

\subsection{Similarity transformation of collective space}
\label{sec3b1}
We begin the discussion of the similarity transformation
(\ref{e114}) by showing that the ${\cal T}$ operator given by,
   \begin{equation}\label{e130}
   {\cal T} = B^i{\cal A}_i
   \end{equation}
leads to the mapping in which bosons and ideal fermions
are entirely decoupled:
   \begin{mathletters}\begin{eqnarray}
   A^j               &\longleftrightarrow&
                            {\cal A}^j - {\textstyle\frac{1}{2}}
                      c^{jl}_{ik}B^i B^k B_l
                                              , \label{e131c} \\
   \big[A_i,A^j\big] &\longleftrightarrow&
                            \big[{\cal A}_i,{\cal A}^j\big]
                    - c^{jl}_{ik}B^kB_l
                                              , \label{e131b} \\
   A_j               &\longleftrightarrow&
                            {\cal A}_j + B_j
                                        , \label{e131a} \\
   a^{\nu}           &\longleftrightarrow&
                            \alpha^{\nu}
                                              , \label{e131e} \\
   a_{\nu}           &\longleftrightarrow&
                            \alpha_{\nu}
                                              , \label{e131d}
   \end{eqnarray}\end{mathletters}
in analogy to the results obtained for the {so(2$N$)}
superalgebra, Eqs.{\ }(\ref{e116c})--(\ref{e116d}).
However, the similarity transformation which is now
responsible for
removing the collective fermion pair ${\cal A}^j$ from the
mapping of $A^j$, Eq.{\ }(\ref{e128c}), is more complicated
than in the corresponding case of the {so(2$N$)} superalgebra.
One has to consider the similarity transformation for
   \begin{equation}\label{e132}
   {\cal T} = \sum_{k=1}^\infty t^{i_1\ldots i_k}_{j_1\ldots j_k}
          {\cal A}^{j_1}\ldots{\cal A}^{j_k}B_{i_1}\ldots B_{i_k} ,
   \end{equation}
where $t^{i_1\ldots i_k}_{j_1\ldots j_k}$ is a totally
symmetric tensor (in upper, as well as in lower indices)
built from the structure constants $c^{jl}_{ik}$. This results
in the mapping
\widetext
   \begin{mathletters}\begin{eqnarray}
   A^j               &\longleftrightarrow&
                          gB^j - {\textstyle\frac{1}{2}}
                    c^{jl}_{ik}B^i B^k B_l
                        - \left(B^i - {\cal A}^k{{\cal T}'}^i_k\right)
                          \left(g\delta^j_i
                    - \big[{\cal A}_i,{\cal A}^j\big]\right)
                                              , \label{e133c} \\
   \big[A_i,A^j\big] &\longleftrightarrow&
                          \big[{\cal A}_i,{\cal A}^j\big]
                    - c^{jl}_{ik}B^kB_l
                                              , \label{e133b} \\
   A_j               &\longleftrightarrow&
                          B_j                 , \label{e133a} \\
   a^{\nu}           &\longleftrightarrow&
                          \alpha^{\nu}
                    + \left(B^i - {\cal A}^k{{\cal T}'}^i_k\right)
                          \left(\big[{\cal A}_i,\alpha^{\nu}\big]
                        - \big[{\cal A}^l,\big[{\cal A}_i,
                    \alpha^{\nu}\big]\big]
                          B_j{{\cal T}'}^j_l\right)
                                              ,  \label{e133e} \\
   a_{\nu}           &\longleftrightarrow&
                          \alpha_{\nu}
                    - \big[{\cal A}^k,\alpha_{\nu}\big]
                          B_i{{\cal T}'}^i_k
                                              ,  \label{e133d}
   \end{eqnarray}\end{mathletters}
provided the operators ${{\cal T}'}^i_j$ and ${{\cal T}''}^{im}_{jn}$,
   \begin{mathletters}\begin{eqnarray}
   {{\cal T}'}^i_j &=&
       \sum_{k=1}^\infty kt^{ii_1\ldots i_{k-1}}_{jj_1\ldots j_{k-1}}
       {\cal A}^{j_1}\ldots{\cal A}^{j_{k-1}}B_{i_1}\ldots B_{i_{k-1}}
                     , \label{e134} \\
   {{\cal T}''}^{im}_{jn} &=&
       \sum_{k=1}^\infty k(k-1)
            t^{imi_1\ldots i_{k-2}}_{jnj_1\ldots j_{k-2}}
       {\cal A}^{j_1}\ldots{\cal A}^{j_{k-2}}B_{i_1}\ldots B_{i_{k-2}}
                     , \label{e135}
   \end{eqnarray}\end{mathletters}
fulfil equations:
   \begin{mathletters}\begin{eqnarray}
   \left(\delta^j_l - g{{\cal T}'}^j_l
      + \frac{1}{2}\left({{\cal T}''}^{kn}_{ml}
      + {{\cal T}' }^k_m{{\cal T}'}^n_l\right)
         c^{ji}_{kn}{\cal A}^m B_i\right){\cal A}^l B_j &=& 0
                                  , \label{e136} \\
   \left(c^{im}_{kl}{{\cal T}'}^j_m
      - c^{ij}_{km}{{\cal T}'}^m_l\right){\cal A}^l B_j &=& 0
                                  . \label{e137}
   \end{eqnarray}\end{mathletters}
\narrowtext

Eqs.{\ }(\ref{e136}) and (\ref{e137}) represent recurrence
relations for tensors $t^{i_1\ldots i_k}_{j_1\ldots j_k}$,
which can be solved for particular structure constants
$c^{jl}_{ik}$. Since the structure constants are not
arbitrary matrices, but obey stringent conditions resulting
from the Jacobi identities for the collective algebra, the
recurrence relations cannot be solved unless these
conditions are properly taken into account. This is
difficult without specifying a particular collective
algebraic structure.  Below we solve the recurrence
relations for the unitary collective algebras.

The intruding term ${\cal A}^i$ is now absent from the mapping
of $A^i$, Eq.{\ }(\ref{e133c}), and
the even-fermion-number collective states are in fact
entirely bosonized. This is so because the last term
in (\ref{e133c}) vanishes when acting on a state
where no ideal fermions exist,
   \begin{equation}\label{e209}
   \left(g\delta^j_i
      - \big[{\cal A}_i,{\cal A}^j\big]\right)|0\rangle = 0 ,
   \end{equation}
cf.{\ }Eq.{\ }(\ref{e60}).

Similarly as for the {so(2$N$)} superalgebra, when an odd
fermion is added to a collective even state,
a series of terms appears in the ideal space.
These terms have one, three, five$\ldots$
ideal bosons added to purely bosonic components.

When the next fermion is added to an odd state
the ideal fermions will not in general disappear from the
corresponding boson-fermion image. This reflects
the fact that when two real arbitrary fermions are added to
a collective even state, this state will not in general
belong to the collective space of the next
even nucleus.
The collective superalgebra closure relations
(\ref{e5x}) do not ensure that the
corresponding supergenerators leave the collective
space invariant. This is obvious for the
single-fermion creation operators $a^{\mu}$,
which create the complete fermion
Fock space and therefore cannot conserve the collective
space.

On the other hand, when two fermions are added to
a collective even state, and the appropriate
linear combination is then taken as in
Eq.{\ }(\ref{e4}), so as to form a collective pair,
the resulting state does belong to the collective
space of the next even nucleus. If an analogous
operation is performed in the ideal space, one
observes the desired mechanism of a recombination
of odd ideal fermions into bosons. More precisely,
by acting on the series of terms which represents
an odd ideal state with the series of terms
(\ref{e140e}) which represents a single
fermion operator, and next forming a collective pair
(\ref{e4}), one sees that the two series conspire in such
a way that
ideal fermions disappear from the resulting expression.

The similarity mapping of the collective
superalgebra faithfully represents properties
of the underlying collective space. One obtains
an exact description of Pauli correlations
between bosons representing collective even states and
an odd fermion and repeated application of the images
Eqs.{\ }(\ref{e131e}) and (\ref{e131c}) onto the ideal
space vacuum will yield the physical subspace described
below Eqs.{\ }(\ref{e5x}).

\section{EXAMPLES OF MAPPINGS \\ FOR COLLECTIVE SPACES}
\label{sec4}

\subsection{Similarity mapping for unitary superalgebra}
\label{sec3b2}
Let us suppose that the collective operators form an
$(\Omega+1)$-dimensional symmetric representation of the
unitary algebra su($l$+1), i.e., one has $l$ collective
pairs $A_i$.  The simplest example is provided by the
well-known quasispin su(2) algebra.  By normalizing the
collective pairs so that $g$=$\Omega$ one obtains
$\Omega$-independent structure constants:
   \begin{equation}\label{e138}
   c^{jl}_{ik} = \delta^j_i\delta^l_k + \delta^j_k\delta^l_i .
   \end{equation}
The recurrence relations can now be fulfilled by requiring
that tensors $t^{i_1\ldots i_k}_{j_1\ldots j_k}$ are
proportional to symmetrized products of the Kronecker
delta's. This is equivalent to postulating the
operator ${\cal T}$ to be a function of the operator ${\cal X}$,
${\cal T} = {\cal T}({\cal X})$, where
   \begin{equation}\label{e139}
   {\cal X} = {\cal A}^i B_i
   \end{equation}
and leads to the following mapping:
\widetext
   \begin{mathletters}\begin{eqnarray}
   A^j               &\longleftrightarrow&
                       B^j(\Omega - N_B)
                     - \left(B^i - {\cal A}^i{{\cal T}'}\right)
                       \left(\Omega\delta^j_i
                 - \big[{\cal A}_i,{\cal A}^j\big]\right)
                                              , \label{e140c} \\
   \big[A_i,A^j\big] &\longleftrightarrow&
                       \delta^j_i(\Omega-N_B) - B^jB_i
                     - \left(\Omega\delta^j_i
                 - \big[{\cal A}_i,{\cal A}^j\big]\right)
                                              , \label{e140b} \\
   A_j               &\longleftrightarrow&
                       B_j                    , \label{e140a} \\
   a^{\nu}           &\longleftrightarrow&
                       \alpha^{\nu}
                 + \left(B^i - {\cal A}^i{{\cal T}'}\right)
                       \left(\big[{\cal A}_i,\alpha^{\nu}\big]
                     - \big[{\cal A}^j,\big[{\cal A}_i,
                   \alpha^{\nu}\big]\big]
                       B_j{{\cal T}'}\right)
                                              , \label{e140e} \\
   a_{\nu}           &\longleftrightarrow&
                       \alpha_{\nu}
                 - \big[{\cal A}^i,\alpha_{\nu}\big]
                       B_i{{\cal T}'}
                                              , \label{e140d}
   \end{eqnarray}\end{mathletters}
where $N_B$=$B^kB_k$ is the boson-number operator.
\narrowtext

In these Equations, the operator
${\cal T}'$ obeys the Ricatti equation \cite{Ricatti}
   \begin{equation}\label{e141}
   {\cal X}({\cal T}'' + {{\cal T}'}^2) - \Omega{\cal T}' + 1 = 0 .
   \end{equation}
Recalling that the number of bosons in the physical space
is limited to $\Omega$ we have that $({\cal X})^{\Omega+1}$=0
and the solution can be postulated in the form of a
polynomial,
   \begin{equation}\label{e142}
   {\cal T}' = \sum_{k=0}^\Omega \lambda'_k {{\cal X}}^k ,
   \end{equation}
with the coefficients $\lambda'_k$ determined from the recurrence
relation
   \FL\begin{equation}\label{e143}
   \lambda'_0 = \frac{1}{\Omega} \quad,\quad
   \lambda'_k = \frac{1}{\Omega-k+1}
          \sum_{m=1}^k \lambda'_{m-1}\lambda'_{k-m} .
   \end{equation}
We see that the $(\Omega+1)$-th coefficient becomes
singular, but this of course does not influence the solution
(\ref{e142}).  One also notes that for large $\Omega$
the series (\ref{e142}) is rapidly converging,
   \begin{equation}\label{e208}
   {\cal T}' = \frac{1}{\Omega}
        + \frac{{\cal X}}{\Omega^3}
        + \frac{2{\cal X}^2}{\Omega^4(\Omega-1)}
        + \ldots
   \end{equation}

\subsection{Similarity mapping for the quasispin \\
                   su(2) superalgebra}
\label{sec3b3}
We conclude this section by specifying
Eqs.{\ }(\ref{e128c})--(\ref{e128d}) and
(\ref{e140c})--(\ref{e140d}) for the simplest case of the
quasispin su(2) algebra composed of the single
pair-creation operator $A^+$=$ \sum_{\mu>0} a^{\mu}
a^{\bar\mu}$, its hermitian conjugate $A$ which is the
pair-annihilation operator, and of the fermion-number
operator $N$.  The boson-fermion mapping of
Eqs.{\ }(\ref{e128c})--(\ref{e128d}) then reads
   \begin{mathletters}\begin{eqnarray}
   A^+     &\longleftrightarrow&
           \Omega B^{\mbox{\footnotesize{\dag}}}
              - B^{\mbox{\footnotesize{\dag}}}
              B^{\mbox{\footnotesize{\dag}}} B
              - B^{\mbox{\footnotesize{\dag}}} {\cal N}
              + {\cal A}^{\mbox{\footnotesize{\dag}}}
                                       , \label{e144c} \\
   N       &\longleftrightarrow&
                 2B^{\mbox{\footnotesize{\dag}}} B + {\cal N}
                                   , \label{e144b} \\
   A       &\longleftrightarrow&
                  B                     , \label{e144a} \\
   a^{\mu} &\longleftrightarrow&
                  \alpha^{\mu}
              + B^{\mbox{\footnotesize{\dag}}} \alpha_{\bar\mu}
                                       , \label{e144e} \\
   a_{\mu} &\longleftrightarrow&
                  \alpha_{\mu}         . \label{e144d}
   \end{eqnarray}\end{mathletters}

The similarity transformation $e^{-{\cal T}}{\cal O} e^{\cal T}$, for
${\cal T}$ given by Eqs.{\ }(\ref{e142}) and (\ref{e143}) and
${\cal X}$=${\cal A}^{\mbox{\footnotesize{\dag}}}B$,
removes the ideal fermion pair from the
physical space of an even system:
   \FL\begin{mathletters}\begin{eqnarray}
   A^+     &\longleftrightarrow&
           \Omega B^{\mbox{\footnotesize{\dag}}}
              - B^{\mbox{\footnotesize{\dag}}}
              B^{\mbox{\footnotesize{\dag}}} B
          - \left(B^{\mbox{\footnotesize{\dag}}} - {\cal T}'
         {\cal A}^{\mbox{\footnotesize{\dag}}}\right){\cal N}
                                       , \label{e145c} \\
   N       &\longleftrightarrow&
                 2B^{\mbox{\footnotesize{\dag}}} B + {\cal N}
                                   , \label{e145b} \\
   A       &\longleftrightarrow&
                  B                    , \label{e145a} \\
   a^{\mu} &\longleftrightarrow&
                  {\cal T}'\left(\Omega
              - B^{\mbox{\footnotesize{\dag}}} B\right)\alpha^{\mu}
          + \left(B^{\mbox{\footnotesize{\dag}}} - {\cal T}'
         {\cal A}^{\mbox{\footnotesize{\dag}}}\right)
                \alpha_{\bar\mu}     , \label{e145e} \\
   a_{\mu} &\longleftrightarrow&
                  \alpha_{\mu} + {\cal T}'\alpha^{\bar\mu}B
                                       . \label{e145d}
   \end{eqnarray}\end{mathletters}
Boson-fermion images of even and odd collective states
have the following form
   \FL\begin{mathletters}\begin{eqnarray}
       (A^+)^N|0\rangle &\longleftrightarrow&
      \frac{\Omega!}{(\Omega-N)!}
        (B^{\mbox{\footnotesize{\dag}}})^N|0) ,
                                          \label{e210a} \\
   a^{\mu}(A^+)^N|0\rangle &\longleftrightarrow&
      \frac{\Omega!}{(\Omega-N-1)!}
         \alpha^{\mu}{\cal T}'(B^{\mbox{\footnotesize{\dag}}})^N|0) ,
                                          \label{e210b}
   \end{eqnarray}\end{mathletters}
where we see specifically that the single ideal fermion states
($N=0$ in Eq.{\ }(\ref{e210b})) are correctly normalized:
   \begin{equation}\label{e210c}
   a^\mu|0\rangle \longleftrightarrow \alpha^\mu|0) \; .
   \end{equation}
For the even non-collective states one finds e.g.
\widetext
   \FL\begin{equation}\label{e211}
   a^{\mu} a^{\bar\mu}(A^+)^N|0\rangle  \longleftrightarrow
      \frac{(\Omega-1)!}{(\Omega-N-1)!}
        (B^{\mbox{\footnotesize{\dag}}})^{N+1}|0)
      + \left[(\Omega-B^{\mbox{\footnotesize{\dag}}}B){\cal T}'^2
      + {\cal T}'\right]
        \left[\Omega\alpha^{\mu}\alpha^{\bar\mu}
      - {\cal A}^{\mbox{\footnotesize{\dag}}}\right]
      \frac{(\Omega-1)!}{(\Omega-N)!}
        (B^{\mbox{\footnotesize{\dag}}})^N    |0) .
   \end{equation}
\narrowtext\noindent
When $\Omega$ non-collective pairs $a^{\mu} a^{\bar\mu}$
are summed together to form the collective pair $A^+$,
the second term in the image (\ref{e211}) vanishes because
${\cal A}^{\mbox{\footnotesize{\dag}}}=
\sum_{\mu>0} \alpha^{\mu}\alpha^{\bar\mu}$.
The resulting image of the even collective
state reduces to the state with $N$+1 collective bosons,
as it should.

The images (5.9) bear a strong resemblance to similar results
obtained in Ref.{\ }\cite{GH83} and the two sets must in fact
be related by a further similarity transformation which we
have so far not been able to find.

\section{DISCUSSION}
\label{sec5}

We have presented a framework which extends the construction
of boson mappings through coherent states to the domain of
boson-fermion mappings. This is accomplished by the introduction
of Grassmann variables into supercoherent states. Calculations
were facilitated by the identification and further use of the
associated Usui operators.

The formalism allowed us to construct a known Dyson-type
mapping for the full {so(2$N$+1)} algebra, together with some
other formal mappings not previously considered. We also
obtained some first results for boson-fermion mappings
relevant to collective subspaces. However, additional effort
will have to be directed at this aspect on two levels.

On the
formal level one may think in terms of solving the
recurrence relations in
Eqs.{\ }(\ref{e136}) and (\ref{e137}) for other examples than
unitary algebras. Since low rank orthogonal algebras have
played a prominent role in fermion models with dynamical symmetry,
these seem to be of most immediate interest. Alternatively,
or additionally, one may need
innovation in either the construction of novel
supercoherent states or appropriate similarity transformations.
Furthermore, utilization of the results to make further
contact between microscopic models and (semi-) phenomenological
models such as the IBFM, is also called for.

\acknowledgments

This work was supported by a grant from the Foundation for
Research Development of South Africa, the University of
Stellenbosch and in part by the Polish State Committee for
Scientific Research under Contract No.  20450~91~01.

\appendix{Boson-fermion mapping of the \\
             {so(2$N$)} superalgebra}
\label{appA}

In order to drive the boson-fermion mapping of the {so(2$N$)}
superalgebra, Eqs.{\ }(\ref{e110c})--(\ref{e110d}), we use
the standard method \cite{Dob81a} of commuting real and
ideal operators with the Usui operator (\ref{e107}).  Let
us denote by ${{\cal U}}$ the exponent appearing in the
definition of the Usui operator,
   \begin{equation}\label{e212}
   {\cal U} = \exp\left({\cal C}\right)
       = \exp\left({\textstyle\frac{1}{2}} B^{\mu\nu}a_{\nu}a_{\mu} +
                  \alpha^{\mu} a_{\mu}\right) ,
   \end{equation}
which acts in the product space of real and ideal states.
We will apply to Eq.{\ }(\ref{e108}) two forms of the BCH
formula (\ref{e114}),
   \begin{mathletters}\begin{eqnarray}
   {\cal O}{\cal U}    &=& {\cal U}\sum_{k=0}^\infty \frac{(-1)^k}{k!}
         [{\cal C}[{\cal C}\ldots[{\cal C},{\cal O}]\ldots]]_k  ,
                                 \label{e213a} \\
   {\cal U}\hat O &=&    \sum_{k=0}^\infty \frac{1}     {k!}
         [{\cal C}[{\cal C}\ldots[{\cal C},\hat O]\ldots]]_k {\cal U},
                                 \label{e213b}
   \end{eqnarray}\end{mathletters}
remembering that after the calculation of multiple
commutators, ${\cal U}$ acts on the ideal (real) vacuum to the
right (left).

We first consider the ideal fermion annihilation operator
and Eq.{\ }(\ref{e213a}),
   \begin{equation}\label{e214}
   \alpha_{\nu}{\cal U} = {\cal U}\left(\alpha_{\nu}
                 - [\alpha^{\mu} a_{\mu},\alpha_{\nu}]\right) .
   \end{equation}
The first term gives zero when acting to the right on the
ideal vacuum, while the commutator reads
   \begin{equation}\label{e215}
   [\alpha^{\mu} a_{\mu},\alpha_{\nu}] = - a_{\nu}
   \end{equation}
(recall that ideal fermions anticommute with real
fermions), whence the higher-order multiple commutators
vanish.  Therefore, one obtains
   \begin{equation}\label{e216}
   \alpha_{\nu} U = U a_{\nu} ,
   \end{equation}
i.e., the mapping (\ref{e110d}) is proved.

Second, we consider the real fermion creation operator
and Eq.{\ }(\ref{e213b}),
   \begin{equation}\label{e217}
   {\cal U} a^{\nu} = \left(a^{\nu}
     + {\textstyle\frac{1}{2}}[B^{\mu\rho} a_{\rho} a_{\mu},a^{\nu}]
     + [\alpha^{\mu} a_{\mu},a^{\nu}]\right) {\cal U} .
   \end{equation}
Again, the first term vanishes when acting to the left on
the real vacuum, and the commutators read
   \begin{equation}\label{e218}
       {\textstyle\frac{1}{2}}[B^{\mu\rho} a_{\rho} a_{\mu},a^{\nu}]
     + [\alpha^{\mu} a_{\mu},a^{\nu}]
     = B^{\nu\rho} a_{\rho} + \alpha^{\nu} .
   \end{equation}
where both terms commute with ${\cal C}$ and ${\cal U}$.  Using the
previously derived Equation (\ref{e216}) one finally has
that
   \begin{equation}\label{e219}
   \left(B^{\nu\rho}\alpha_{\rho} + \alpha^{\nu}\right) U = U a^{\nu},
   \end{equation}
i.e., the mapping (\ref{e110e}) is proved.

Continuing similar derivations, one may consider $B_{\mu\nu}{\cal U}$
to prove mapping (\ref{e110a}), then ${\cal U} a^{\mu} a_{\nu}$ to
prove (\ref{e110b}), and finally ${\cal U} a^{\mu} a^{\nu}$ to prove
(\ref{e110c}).

\appendix{Similarity transformations in \\
          the ideal boson-fermion space}
\label{appB}

In order to derive the similarity-transformed boson-fermion
images, Eqs.{\ }(\ref{e116c})--(\ref{e116d}) and
Eqs.{\ }(\ref{e119c})--(\ref{e119d}), one first considers
the multiple commutators in the BCH formula,
Eq.{\ }(\ref{e114}), where the operator ${\cal T}$ is given as a
power series (\ref{e120}) of ${\cal X}$.  We will only consider
such operators ${\cal X}$ and ${\cal O}$ that
   \begin{equation}\label{e221}
   [{\cal X},[{\cal X},{\cal O}]] = 0  .
   \end{equation}
In this case the commutator acts on a power series like a
differentiation, i.e.
   \begin{equation}\label{e222}
   [{\cal X}^k,{\cal O}] = k{\cal X}^{k-1}[{\cal X},{\cal O}]
   \end{equation}
and
   \begin{equation}\label{e223}
   [{\cal T},{\cal O}] = {\cal T}'[{\cal X},{\cal O}] .
   \end{equation}
Moreover, the multiple commutators vanish,
   \begin{equation}\label{e224}
   [{\cal T},[{\cal T},{\cal O}]] = 0 ,
   \end{equation}
and the BCH formula reduces to
   \begin{equation}\label{e225}
   {\cal W}^{-1}{\cal O}{\cal W} =
   {\cal O} - {\cal T}'[{\cal X},{\cal O}] .
   \end{equation}

For the operator ${\cal T}$ given by Eq.{\ }(\ref{e115}) one
therefore obtains the following similarity transformations:
   \begin{equation}\begin{array}{rcl}\label{e226}
   {\cal W}^{-1}B^{\mu\nu}  {\cal W} &=& B^{\mu\nu}               , \\
   {\cal W}^{-1}B_{\mu\nu}  {\cal W} &=& B_{\mu\nu}
                                       + \alpha_{\nu}\alpha_{\mu} , \\
   {\cal W}^{-1}\alpha^{\nu}{\cal W} &=& \alpha^{\nu}
                                       - B^{\nu\rho}\alpha_{\rho} , \\
   {\cal W}^{-1}\alpha_{\nu}{\cal W} &=& \alpha_{\nu}             , \\
   \end{array}\end{equation}
which applied to the boson-fermion images in
Eqs.{\ }(\ref{e110c})--(\ref{e110d}) give those in
Eqs.{\ }(\ref{e116c})--(\ref{e116d}).  Note that some
products of ideal operators, like $\alpha^{\mu}\alpha^{\nu}$ for
example, do not fulfil condition (\ref{e221}).  Their
similarity transformation can, however, be calculated as
products of transformations of separate factors, which do
fulfil (\ref{e221}).

When ${\cal T}$ is given as a power series in ${\cal X}$,
Eq.{\ }(\ref{e117}), one has
   \begin{equation}\begin{array}{rcl}\label{e227}
   {\cal W}^{-1}B^{\mu\nu}  {\cal W} &=& B^{\mu\nu}
                          - {\cal T}'\alpha^{\mu}\alpha^{\nu}  , \\
   {\cal W}^{-1}B_{\mu\nu}  {\cal W} &=& B_{\mu\nu}            , \\
   {\cal W}^{-1}\alpha^{\nu}{\cal W} &=& \alpha^{\nu}          , \\
   {\cal W}^{-1}\alpha_{\nu}{\cal W} &=& \alpha_{\nu}
                          - {\cal T}'B_{\mu\nu}\alpha^{\mu}    , \\
   \end{array}\end{equation}
and the mapping in Eqs.{\ }(\ref{e119c})--(\ref{e119d}) is
obtained by inserting these similarity transformations in
Eqs.{\ }(\ref{e110c})--(\ref{e110d}).  For example, the
similarity image of the single-fermion creation operator
reads
   \FL\begin{equation}\label{e228}
   a^{\nu} \longleftrightarrow \alpha^{\nu}
                  + \left(B^{\nu\rho}
                - {\cal T}'\alpha^{\nu}\alpha^{\rho}\right)
                    \left(\alpha_{\rho}
                  - {\cal T}'B_{\theta\rho}\alpha^{\theta}\right) .
   \end{equation}
After normal-ordering and grouping together terms with
$\alpha^{\nu}$ one obtains
   \FL\begin{eqnarray}
    a^{\nu} &\longleftrightarrow&
                \alpha^{\nu}\left(1-{\cal T}'{\cal N}
              - ({\cal T}''+{\cal T}'^2)\alpha^{\rho}\alpha^{\theta}
                B_{\rho\theta}\right)
                               \nonumber   \\
       &    &   \phantom{\alpha^{\nu}(1-{\cal T}'{\cal N}-}
              - \alpha^{\theta}{\cal T}'B^{\nu\rho} B_{\theta\rho}
            + B^{\nu\rho}\alpha_{\rho} .
                               \label{e229}
   \end{eqnarray}
The term with second derivative ${\cal T}''$ appears as a result
of commuting $B^{\nu\rho}$ and ${\cal T}'$.  After using the Ricatti
Equation (\ref{e121}) one obtains mapping (\ref{e119e}).

\appendix{Boson-fermion mapping of the \\
            {so(2$N$)} superalgebra using  \\
          projected supercoherent states}
\label{appC}

Similarly as in Eq.{\ }(\ref{e212}), we define the
projected Usui operator in the product space as
   \begin{equation}\label{e230}
   {\cal U}_{01} = \exp({\cal C})(1+\alpha^{\mu} a_{\mu})
            =     (1+\alpha^{\mu} a_{\mu})\exp({\cal C})
   \end{equation}
for
   \begin{equation}\label{e231}
   {\cal C} = {\textstyle\frac{1}{2}} B^{\mu\nu} a_{\nu} a_{\mu} .
   \end{equation}
Then we use the BCH formula to show that
   \begin{equation}\label{e232}
   a_{\mu} a_{\nu} \exp({\cal C}) = B_{\nu\mu}\exp({\cal C}) .
   \end{equation}
Considering the fermion annihilation operator one has
   \begin{equation}\label{e233}
   {\cal U}_{01}a_{\nu} =
      (a_{\nu} + \alpha^{\mu} a_{\mu} a_{\nu})\exp({\cal C})
   \end{equation}
and the pair of fermions in the second term can be replaced
by a boson as in Eq.{\ }(\ref{e232}), while the first term,
when acting on the ideal vacuum, can be replaced by an
ideal fermion, i.e.,
   \FL\begin{equation}\label{e234}
   {\cal U}_{01}a_{\nu}|0) =
       \alpha_{\nu}(1+\alpha^{\mu}  a_{   \mu})\exp{{\cal C}}|0) +
                      \alpha^{\mu}  B_{\nu\mu} \exp{{\cal C}}|0) .
   \end{equation}
In order to obtain $U_{01}$ in the second term on the
right-hand side, we need to use the projection operator
${\cal Q}$, Eq.{\ }(\ref{e205}), which conserves the ideal
vacuum and annihilates one-ideal-fermion states.  Then one
obtains
   \begin{equation}\label{e235}
   U_{01}a_{\nu} = (\alpha_{\nu}
       + \alpha^{\mu} B_{\nu\mu}{\cal Q})U_{01} ,
   \end{equation}
and mapping (\ref{e113d}) is proved.

Similarly, we use the BCH formula to show that
   \begin{equation}\label{e236}
   \exp({\cal C})a^{\nu} =
       (a^{\nu} + B^{\nu\rho} a_{\rho})\exp({\cal C}) .
   \end{equation}
Considering the fermion creation operator one therefore has
   \begin{equation}\label{e237}
   {\cal U}_{01}a^{\nu} =
     (1+\alpha^{\mu} a_{\mu})(a^{\nu}
        + B^{\nu\rho} a_{\rho})\exp({\cal C})
   \end{equation}
and
   \begin{equation}\label{e238}
   \langle0|{\cal U}_{01}a^{\nu} =
      \langle0|\alpha^{\nu}\exp({\cal C}) +
             B^{\nu\rho}\langle0|{\cal U}_{01}a_{\rho} .
   \end{equation}
Eq.{\ }(\ref{e235}) can now be used to transform the second
term, while the ${\cal Q}$ operator is again necessary to obtain
$U_{01}$ in the first term.  Finally one obtains
   \FL\begin{equation}\label{e239}
   U_{01}a^{\nu} = \alpha^{\nu}{\cal Q} U_{01}
         + (B^{\nu\rho}\alpha_{\rho}
       +  B^{\nu\rho}\alpha^{\mu} B_{\nu\mu}{\cal Q})U_{01} ,
   \end{equation}
and mapping (\ref{e113e}) is proved.

One notes that the possibility to replace in
Eq.{\ }(\ref{e233}) an arbitrary fermion-pair annihilation
operator by a boson-annihilation operator is the key
element of the derivation.  When considering collective
algebras such a replacement is not possible, and therefore
a projected mapping cannot be similarly derived in the
collective space.

We conclude this appendix with an example of
how functional images
are directly utilized to derive operator mappings.
The mapping (\ref{e113d}) is derived in this manner by
defining a supercoherent state projected to a space with
zero or one ideal fermions
\begin{eqnarray}
 \label{e240}
\langle{C,\phi}| &:=     & \langle{0}|(1+\phi_\nu a_\nu)
                           \exp({\textstyle\frac{1}{2}}
                           C_{\mu\nu}a_\nu a_\mu) \nonumber \\
                 &\equiv & \langle{0}|(1+\phi_\nu a_\nu)
                       e^{\hat C}\; ,
\end{eqnarray}
similar to the state (\ref{e102}), except for the projection.

The image of $a_\mu$ relevant to the above space is now
constructed as follows.
\begin{eqnarray}\label{e241}
   \langle{C,\phi}|a_\mu
        &=& \langle{0}|(1+\phi_\nu a_\nu)a_\mu
                 e^{\hat C}\nonumber\\
        &=& \langle{0}|(a_\mu+\phi_\nu a_\nu a_\mu)
                 e^{\hat C}\nonumber\\
        &=& \langle{0}|(a_\mu+\phi_\nu \partial_{\mu\nu)}
                 e^{\hat C}\nonumber\\
        &=& \partial_\mu \langle{0}|(1+\phi_\mu a_\mu)
                 e^{\hat C} + \langle{0}| \phi_\nu \partial_{\mu\nu)}
                 e^{\hat C} \nonumber\\
        &=& (\partial_\mu+\phi_\nu\partial_{\mu\nu}{\cal Q})
            \langle{0}|(1+\phi_\mu a_\mu)e^{\hat C} \nonumber\\
        &=& (\partial_\mu+\phi_\nu\partial_{\mu\nu}{\cal Q})
            \langle{C,\phi}| .
\end{eqnarray}

In the second last line it is clear that the projector ${\cal Q}$
must enter in order to extract the supercoherent state
required for the final operator association. This association
is the standard one, namely that a Grassmann variable and its
derivative are associated with, respectively, a (ideal) fermion
creation and a (ideal) fermion annihilation operator, while the
usual Bargmann representation for complex variables is used. From
the (over-)completeness of coherent states one can now clearly
extract from the result (\ref{e241}) the {\it operator} equivalence
(mapping) (\ref{e113d}).

\end{document}